# Imaging antiferromagnetic domain fluctuations and the effect of atomic-scale disorder in a doped spin-orbit Mott insulator

He Zhao[1], Zach Porter[2], Xiang Chen[2], Stephen D. Wilson[2], Ziqiang Wang[1] and Ilija Zeljkovic[1]

[1] *Department of Physics, Boston College, Chestnut Hill, MA 02467, USA*
[2] *Materials Department, University of California Santa Barbara, Santa Barbara, California 93106, USA*

*Correspondence: ilija.zeljkovic@bc.edu

## Abstract

Correlated oxides can exhibit complex magnetic patterns, characterized by domains with vastly different size, shape and magnetic moment spanning the material. Understanding how magnetic domains form in the presence of chemical disorder and their robustness to temperature variations has been of particular interest, but atomic-scale insight into this problem has been limited. We use spin-polarized scanning tunneling microscopy to image the evolution of spin-resolved modulations originating from antiferromagnetic (AF) ordering in a spin-orbit Mott insulator $Sr_3Ir_2O_7$ as a function of chemical composition and temperature. We find that replacing only several percent of La for Sr leaves behind nanometer-scale AF puddles clustering away from La substitutions preferentially located in the middle SrO layer within the unit cell. Thermal erasure and re-entry into the low-temperature ground state leads to a spatial reorganization of the AF modulations, indicating multiple stable AF configurations at low temperature. Interestingly, regardless of this rearrangement, the AF puddles maintain scale-invariant fractal geometry in each configuration. Our experiments reveal spatial fluctuations of the AF order in electron doped $Sr_3Ir_2O_7$, and shed light on its sensitivity to different types of atomic-scale disorder.

## Main Text

Competing energy scales in correlated electron systems often lead to granular nature [1–7], with substantial spatial variations in charge, spin and orbital degrees of freedom. Depending on the material and the emergent property, the characteristic length scale of this inhomogeneity can span several orders of magnitude. For instance, charge inhomogeneity ranges from several micrometers in manganites [2] and Hg-based cuprates [3], to a few nanometers in Bi-based cuprates [1] and iridates [4,5]. Spin inhomogeneity in correlated quantum solids can exhibit an even more elaborate texture. A ferromagnet can have microscopic domains, each with a different orientation of the local magnetic moment [6], sometimes coexisting with spatially separated non-magnetic regions [7]. In the presence of antisymmetric spin interactions, spins can gradually twist and form complex, long wavelength spin textures such as skyrmion lattices [8–10]. While a larger spatial extent of ferromagnetic domains and skyrmion spin texture facilitates imaging by a variety of probes, antiferromagnetic (AF) ordering has been notoriously difficult to visualize.

This is in large part due to the cancellation of signals from neighboring spins, which limits the use of probes that are unable to achieve sub-nanometer resolution. Tremendous progress has recently been made in imaging micron size AF domains using scattering techniques [11]. However, to fully understand the nature of the strongly correlated states and quantum fluctuations, it is highly desirable to visualize the formation of AF clusters with atomic resolution, including their size, shape and distribution in the presence of atomic defects and across different phase transitions.

Over the past decade, Ruddlesden-Popper iridates [12–15] have emerged as an intriguing family of 2D AF spin-orbit Mott insulators, where the AF Mott ground state can be tuned by different parameters, for example temperature and chemical composition. The experiments thus far revealed a remarkable nanoscale phase separation in iridates, reflected in the granular fabric of both their electronic and magnetic properties [4,5,16]. While in many correlated oxides the competition between different phases provides a natural tendency towards phase separation [5,17,18], chemical disorder that is often necessary to induce a new phenomenon can also play a significant role [5,19,20]. Our recent spin-polarized scanning tunneling microscopy (SP-STM) experiments revealed the emergence of AF puddles near insulator-to-metal transition in the Ruddlesden-Popper iridate $Sr_2IrO_4$ [16]. However, many questions remain regarding the formation of AF clusters in this family of oxides. For example, how sensitive is the domain size and shape to the inevitable presence of chemical disorder? How repeatable is the AF pattern configuration after it is thermally wiped out by increasing the temperature, and then cooled back down? In this work, we explore these questions in the bilayer iridate $Sr_3Ir_2O_7$ (Ir-327). Ir-327 is less insulating compared to its single-layer counterpart $Sr_2IrO_4$, which facilities tunneling measurements across a wider doping range down to the near "parent" state. Using SP-STM, we discover that multiple AF configurations can nucleate at low temperature, and reveal the effects of different types of disorder on the local distribution of AF ordering.

Undoped parent Ir-327 exhibits collinear AF ordering below $T_N \sim 285K$, with local magnetic moment of the Ir atoms pointing out-of-plane (Fig. 1(a)) [21,22]. Based on the combination of magnetization and neutron scattering measurements, the long-range AF order can be suppressed by temperature or by chemical substitutions [23,24] (Fig. 2(a)). To access local spin information related to the AF order in this system, we apply SP-STM imaging (Methods, Figure S1) [25]. We acquire a pair of STM topographs over an identical area of the sample in magnetic field applied parallel and antiparallel to the **c**-axis (Fig. 1(c,d)). The reversal of the magnetic field direction serves to "flip" the spin-polarization of the STM tip, without significantly affecting the magnetic moment orientation in the sample. By subtracting the two topographs, we can extract the spin-resolved magnetic contrast map $M(\mathbf{r})$, which shows prominent modulations (Fig. 1(f)). The wave vectors of these modulations, $\mathbf{Q_a} = (0.5, -0.5)$ and $\mathbf{Q_b} = (0.5, 0.5)$ (we hereafter define reciprocal lattice vector $2\pi/a_0 = 1$), are identical to those attributed to long-range AF order by neutron scattering experiments [23,24].

Using this procedure, we investigate the evolution of spin-resolved modulations in Ir-327 as a function of La substituting for Sr [24,26] (Fig. 2(a-c)), while at the same time tracking the electronic properties from differential conductance d$I$/d$V$ spectra (Fig. 2(d-e)). While in the near parent state, we find that the spin-resolved modulations appear spatially uniform (Fig. 2b), in $x$~3.4% La-substituted sample, we reveal a spatial inhomogeneity in the $M(\mathbf{r})$ maps, with distinct regions where modulations are completely absent (Fig. 2c). This spatial inhomogeneity is also reflected in the larger width of the Fourier-space peaks at $Q_a$ and $Q_b$ (Figure S1). The inhomogeneous disappearance of the AF order with La substitution in the bilayer iridate Ir-327 is consistent with our observations in La-substituted $Sr_2IrO_4$ [16], suggesting that the granular nature of the AF order near the transition is not limited to a single iridate. Moreover, out-of-plane spins in Ir-327 allow a complete mapping of all AF domains, in contrast to $Sr_2IrO_4$ where in-plane spin orientation may lead to some domains being undetectable by a spin-polarized STM tip (Methods).

Next, we set to explore if and how chemical disorder affects the AF texture observed after introducing a dilute concentration of La substitutions. We focus on the 3 types of defects we most commonly observe in this system: La substitution for Sr in the topmost SrO plane seen as bright squares in STM topographs ($La_1$), La substitution for Sr in the second SrO layer seen as bright features in d$I$/d$V$(**r**, V) maps at low energy [27] ($La_2$) and dark features in STM topographs ($D_3$) (Fig. 1(a), Fig. 3(a) and Figs. S3, S4). To visualize how the AF order arranges itself relative to local disorder, we superimpose the positions of each type of defect on top of the local AF amplitude map |M(**r**)| (Fig. 3(c-e)). Visually, the AF order appears to be the strongest away from $La_2$ substitutions, while $La_1$ and $D_3$ defects are nearly randomly distributed with respect to the AF order. This conclusion is supported by the cross-correlation of the |M(**r**)| maps and dopant density maps $\rho(r)$ (Fig. 3f, Fig. S4), which shows a cross-correlation coefficient α ~ -0.3 for $La_2$, while α coefficients for $La_1$ and $D_3$ defects are significantly weaker. Bulk measurements of La-doped Ir-327 found that long-range AF order is gradually suppressed with the increase in La concentration [24]. Our SP-STM experiments are consistent with this global picture, and further reveal that the inhomogeneous AF order near the transition point has a tendency to weaken in proximity to a particular type of La disorder, located in the middle SrO plane.

In general, strong disorder pinning could suggest that a spatial pattern associated with the AF order will be repeatable after cycling through the magnetic transition and back. To investigate this, we image the evolution of spin-resolved AF modulations as a function of temperature, tracking the same area of the sample (Fig. 4(a-d)). We find that the average |M(**r**)| signal quickly subsides by ~9 K (Fig. 4f). This provides further evidence that our sample is indeed near the transition point at this La composition. We also note that the spectral gapmaps at 4.5 K and 9 K are nearly identical (Fig. 4(g,h)), thus confirming a lack of direct dependence of the most prominent spectral gap to the local AF order [16]. After increasing the temperature to ~10 K, we cool the sample back down to ~4.5 K and repeat the measurement. Interestingly, although the cross-correlation between the AF amplitude maps before (|$M_1$(5 K)|) and after thermal cycling (|$M_2$(5 K)|) is relatively high (inset in Fig. 4(f)), we observe distinct changes in the patterns (Fig. 4(e)), indicating a change in the low-temperature AF configuration. In both cases, however, the AF patterns form away from $La_2$ substitutions, while other defects play a minor role (Fig. 3f).

The ability to visualize the AF clusters in real-space allows us to investigate their size, shape and distribution in more detail. We apply 2D percolation theory, a powerful statistical analysis tool that can be used to probe the strength of electronic correlations and near-critical behavior from the geometric metrics of the clusters [11,28–30]. We first binarize the AF amplitude map |$M_1$(5K)| based on an intensity cutoff (Fig. 5a, Supplementary Information 2). Then, using logarithmic binning, a standard technique for power-law distribution analysis [31], we plot the relationship between several geometrical descriptors of the clusters: area $A$, perimeter $P$, gyration radius $R_g$ (defined as $2R_g^2 = \sum_{i,j} \frac{|r_i - r_j|^2}{s^2}$, where $r_{i(j)}$ is the position of the $i(j)$-th site in the cluster [31] and $s$ is the size of the cluster), number of clusters of given area $D(A)$ and pair connectivity $PC(r)$ (defined as the probability that two sites separated by a distance $r$ belong to the same connected finite cluster). In the analysis, we exclude the clusters touching the boundaries, as we do not have a complete set of geometric metrics for these. Remarkably, we find that the domain area distribution histogram $D(A)$ vs. $A$ shows a linear dependence on a logarithmic scale spanning over 2.5 decades. By fitting $D(A) = A^{-\tau}$, we determine the critical exponent $\tau = 0.88 \pm 0.24$ (Fig. 5b). Next, we study the relationship between the area $A$ (perimeter $P$) and the gyration radius $R_g$ across all domains, shown in Fig. 5c (Fig. 5d). We find that both $A$ and $P$ scale as a power of $R_g$ ($P = R_g^{dh^*}$ and $A = R_g^{dv^*}$) for exponents $dv^*$ ($dh^*$) often referred to as effective volume (hull) dimensions [31]. From the fitting analysis, we determine these to be $dv^* = 2.0 \pm 0.09$ and $dh^* = 1.20 \pm 0.05$. To put these values in perspective, we note that coefficients for an uncorrelated percolation should be $dv^* \approx 1.90$ and $dh^* = 1.75$ [31]. The difference between these coefficients and those extracted from our data suggest that the formation of AF patterns in our sample is dominated by electronic correlations in the presence of disorder (Supplementary Information 2). To further support the validity of our percolation analysis, we point to the following. First, the exponent ratio $dh^*/dv^*$ determined directly from the slope in Fig. 5e ($3/5 \pm 0.04$) is nearly identical to the ratio of $dh^*$ and $dv^*$ calculated individually from Figs. 5(c,d). Second, we note that the hyperscaling relation, $d - 2 + \eta = 2(d - dv^*)$, expected to hold near the transition point within this model, is satisfied (Supplementary Information 2). Here, $d$ is the dimensionality of the system taken to be 2 given the quasi 2D nature of Ir-327, and $\eta = 0.02 \pm 0.01$ is the coefficient obtained from fitting pair connectivity function $PC(r) \sim r^{-\eta} \cdot e^{-r/\xi}$ (Fig. 5f).

Our experiments provide a new insight into the inhomogeneous distribution of spin degrees of freedom relative to chemical disorder, complementary to the reports of nanoscale separation in the charge degrees of freedom in the same family of materials [4,23,27]. Thermal cycling experiments reveal spatial fluctuations and rearrangement of the AF puddles, which remain positioned largely away from chemical disorder. While our work points towards $La_2$ substitutions in the middle SrO layer having the strongest correlation with the regions where the AF order is absent, it remains to be seen what is the physical effect driving this behavior. The site-dependent effect of La substitutions on the local distribution of the AF order is particularly puzzling, and warrants further experimental and theoretical work. Statistical analysis of the cluster geometric metrics suggests that electronic correlations underpin the domain formation, possibly indicative of near critical behavior in this system. In principle, critical fluctuations can occur on any temporal or spatial length scales. While in this work we reveal spatial variations of the AF order parameter, future experiments could attempt to detect AF fluctuations in time-domain by measuring step-like jumps in tunneling current in constant-height STM mode, in analogy to sudden changes in the optical response in 2D magnets [32]. By exploring the evolution of the coherence length with temperature in the statistical analysis of fractal patterns, future experiments focusing on large sample areas to gain higher statistics could shed light on the underlying criticality in this family of iridates.

**Methods:**

*Single-crystal growth*

Single crystals of $(Sr_{1-x}La_y)_3Ir_2O_7$ were synthesized via a flux growth technique. High-purity powders of $SrCO_3$, $IrO_2$ and $La_2O_3$ (Alfa Aesar) were dried, and stoichiometric amounts were measured out, in a 15:1 molar ratio between $SrCl_2$ flux and the target composition. Powders were loaded into a platinum crucible and further contained inside alumina crucibles to limit volatility. Mixtures were heated to 1300 °C, slowly cooled to 850 °C at a rate of 3.5 °C h$^{-1}$, and then furnace cooled to room temperature. The resulting boule was etched with deionized water, revealing black plate-like crystals with typical dimensions of ~ 1 mm × 1 mm × 0.1 mm. More detailed chemical, structural and electronic characterization can be found in Ref. [24].

*SP-STM measurements*

$(Sr_{1-x}La_x)_3Ir_2O_7$ samples were cleaved in ultrahigh vacuum at ~ 80 K and immediately inserted into the STM head. All STM data were acquired by using a customized Unisoku USM1300 system at the base temperature of ~5 K, unless otherwise noted. A typical STM topograph of reveals a square lattice of Sr atoms in the topmost SrO plane, with the lattice constant $a_0$ ~ 0.39 Å (Fig. 1(c,d)). To produce the spin-resolved magnetic contrast $M(\mathbf{r})$ maps, the Lawler–Fujita drift-correction algorithm [33] was applied to eliminate the inevitable effects of piezo hysteresis and thermal drift, and align the topographic images acquired at two different fields by using various defects as position markers. Spectroscopic measurements were taken using a standard lock-in technique at 915 Hz frequency and varying bias excitation as detailed in the figure captions.

Spin-polarized tips were electrochemically etched from a Cr wire in 2 mol/L NaOH solution, similar to the process described in our previous work [16]. The tip was trained to be ferromagnetic on the surface of ultrahigh-vacuum-cleaved single crystals of antiferromagnet $Fe_{1+y}Te$ by fast scanning and bias pulsing to produce the desired sharpness and the spin-resolved contrast. An advantage of a spin-polarized tip prepared in this method is low "stray" magnetic field because the bulk of the tip remains antiferromagnetic. Lastly, we note that in the single-layer iridate $Sr_2IrO_4$, spins are oriented in-plane, which can lead to AF domains with spins rotated by 90 degrees in-plane [16]. This can in principle lead to some AF domains that are undetectable by an STM tip, if tip polarization is exactly orthogonal to the spin orientation of the sample (see more details in Supplementary Information of Ref. 16). This is not the case in $Sr_3Ir_2O_7$ because spins point out-of-plane and no orthogonally oriented domains should exist.

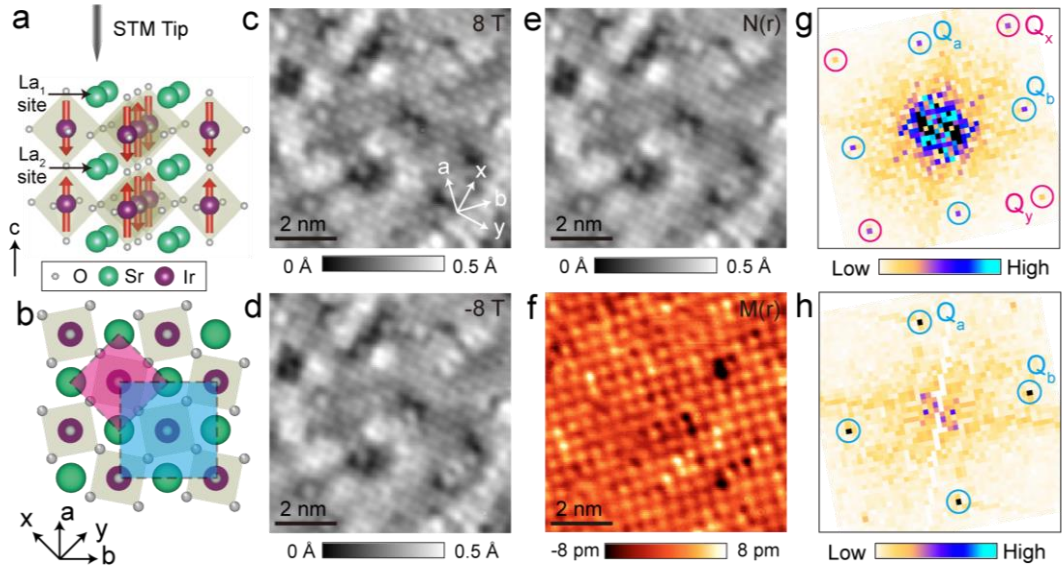

**Figure 1.** Schematic of the crystal structure and the magnetic structure of $Sr_3Ir_2O_7$ and spin-polarized scanning tunneling microscopy imaging. (**a**) 3D crystal structure and (**b**) *ab*-plane atomic structure of $Sr_3Ir_2O_7$. Red arrows in (a) denote the direction of Ir spin moment. The two inequivalent La substitution sites are marked by black arrows in (a): $La_1$ (substitution for Sr in the top or bottom SrO plane) and $La_2$ (substitution for Sr in the middle SrO plane). Two unit cells are outlined by squares in (**b**): the Sr lattice unit cell (light pink) and the magnetic unit cell (light blue). (**c,d**) STM topographs $T(\mathbf{r}, \mathbf{B})$ of near-parent $Sr_3Ir_2O_7$ acquired using a spin-polarized tip, in magnetic field of $\mathbf{B}$ = +8T (**c**) and $\mathbf{B}$ = -8T (**d**) applied perpendicular to the sample surface (+(-) sign denotes the field parallel (antiparrallel) to **z**). (**e**) $N(\mathbf{r})$ map, defined as the arithmatic average of STM topographs in (**c**) and (**d**). (**f**) Spin-resolved magnetic contrast $M(\mathbf{r})$ map, obtained by the subtraction of STM topographs in (**c**) and (**d**), box car averaged with a 1 pixel radius. (**g,h**) Fourier transform of $N(\mathbf{r})$ and $M(\mathbf{r})$ maps shown in (**e**) and (**f**), respectively. $\mathbf{Q_x}$ and $\mathbf{Q_y}$ represent Sr lattice atomic Bragg peaks, while $\mathbf{Q_a}$ and $\mathbf{Q_b}$ denote the Fourier peaks related to spin-resolved AF modulations. STM setup condition: $V_{sample}$ = 1 V, $I_{set}$ = 30 pA, $\mathbf{B}$ = +/-8 T.

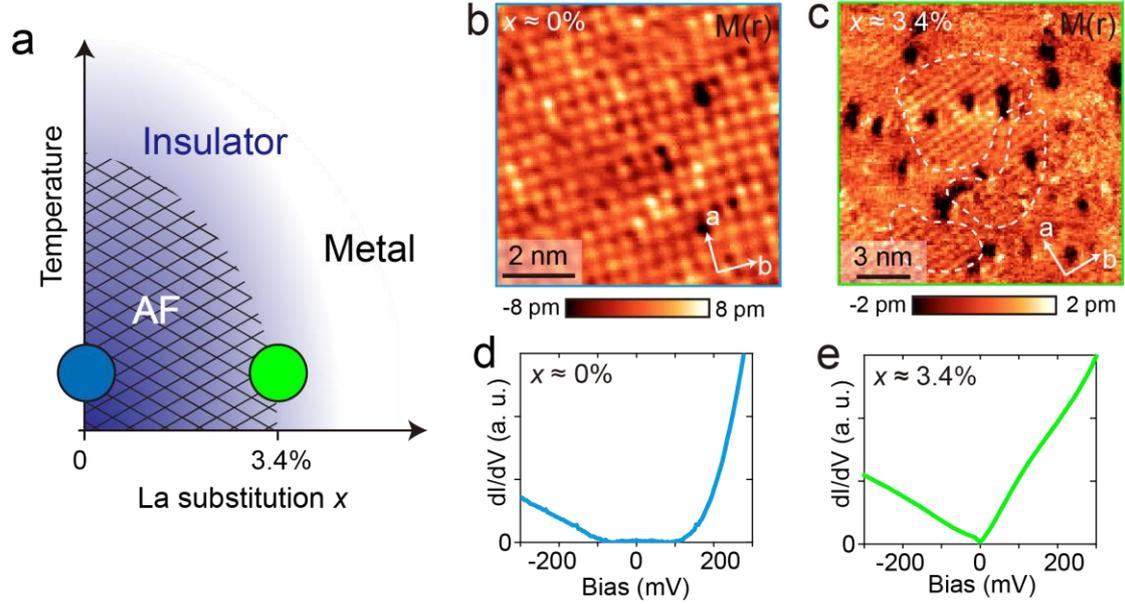

**Figure 2.** Evolution of spin resolved modulations and differential conductance as a function of composition in $(La_xSr_{1-x})_3Ir_2O_7$. (**a**) Schematic phase diagram of $(La_xSr_{1-x})_3Ir_2O_7$ as a function of La concentration $x$. The squared mesh roughly outlines the antiferromagnetic order phase detected by magnetization measurements and scattering experiments [23,24]. The purple (white) background represent the insulator (metal) phase. Two different compositions measured in this work are: near parent $x \sim 0\%$ (blue circle), and $x \sim 3.4\%$ (green circle). (**b-c**) Spin resolved $M(r)$ maps (box car averaged by 1 pixel radius), and (**d-e**) associated spatially averaged $dI/dV(r,V)$ spectra for (**b,d**) $x \sim 0\%$ and (**c,e**) $x \sim 3.4\%$ sample. STM setup condition: (**b**) $V_{sample} = 1$ V, $I_{set} = 30$ pA, +/- 8 T; (**c**) $V_{sample} = 600$ mV, $I_{set} = 100$ pA, +/- 4 T; (**d**) $V_{sample} = 400$ mV, $I_{set} = 100$ pA, $V_{exc} = 4$ mV (zero-to-peak), 0 T; (**e**) $V_{sample} = 300$ mV, $I_{set} = 300$ pA, $V_{exc} = 6$ mV (zero-to-peak), 0 T.

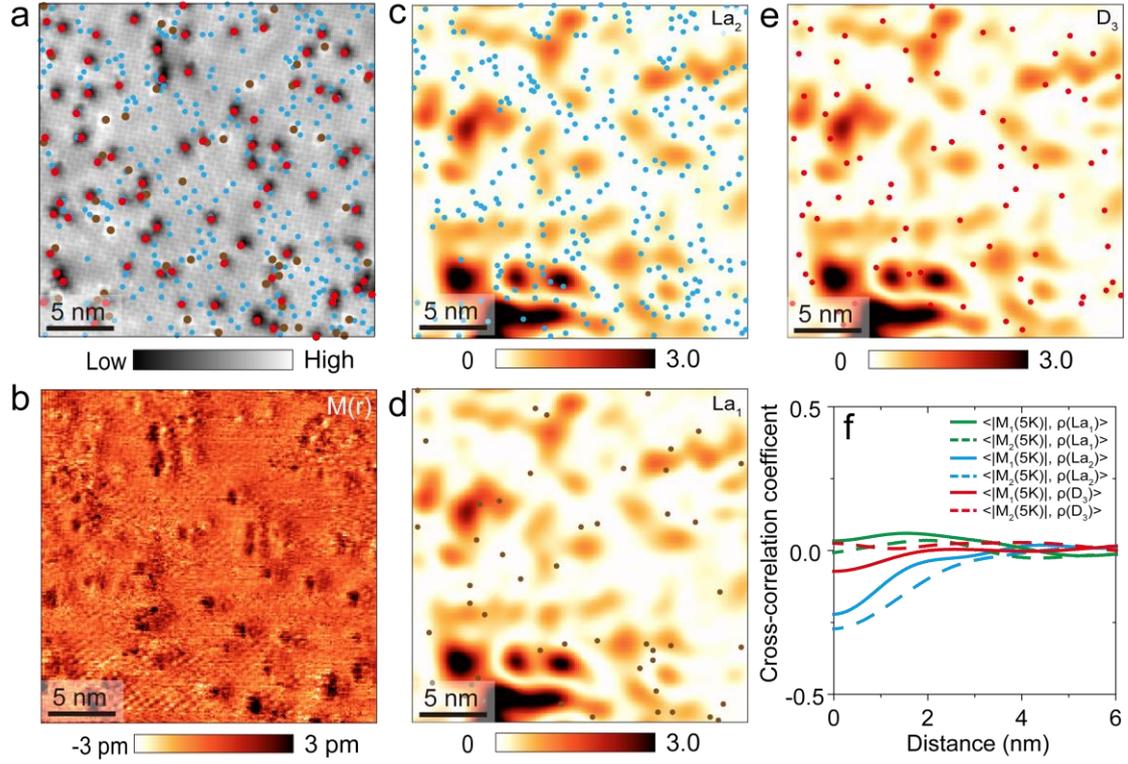

**Figure 3.** Cross-correlation between spin-resolved magnetic contrast map $M(\mathbf{r})$ and defect distribution in $x\sim3.4\%$ $Sr_3Ir_2O_7$ sample. (**a,b**) $N(\mathbf{r})$ ($M(\mathbf{r})$) map, defined as the arithmetic average (subtraction) of two topographs $T(\mathbf{r}, \mathbf{B})$ acquired using a spin-polarized tip with different spin polarizations. The positions of three different types of dopants are denoted by circles of different color: $La_1$ (substitution for Sr in the top SrO plane) (brown), $La_2$ (substitution for Sr in the middle SrO plane) (light blue), and another unidentified defect labeled $D_3$. (**c-e**) $M(\mathbf{r})$ amplitude map $|M_1(\mathbf{r}, 5K)|$ with positions of different defects superimposed: $La_2$ (**c**), $La_1$ (**d**), and $D_3$ (**e**). (**f**) Radially-averaged cross-correlation between $M(\mathbf{r})$ amplitude maps at 5 K (initial cooldown $|M_1(\mathbf{r}, 5K)|$ and after thermal cycling $|M_2(\mathbf{r}, 5K)|$) and dopant density maps (Fig. S4). STM setup conditions: (a,b) $V_{sample} = 600$ mV, $I_{set} = 100$ pA, +/- 4 T.

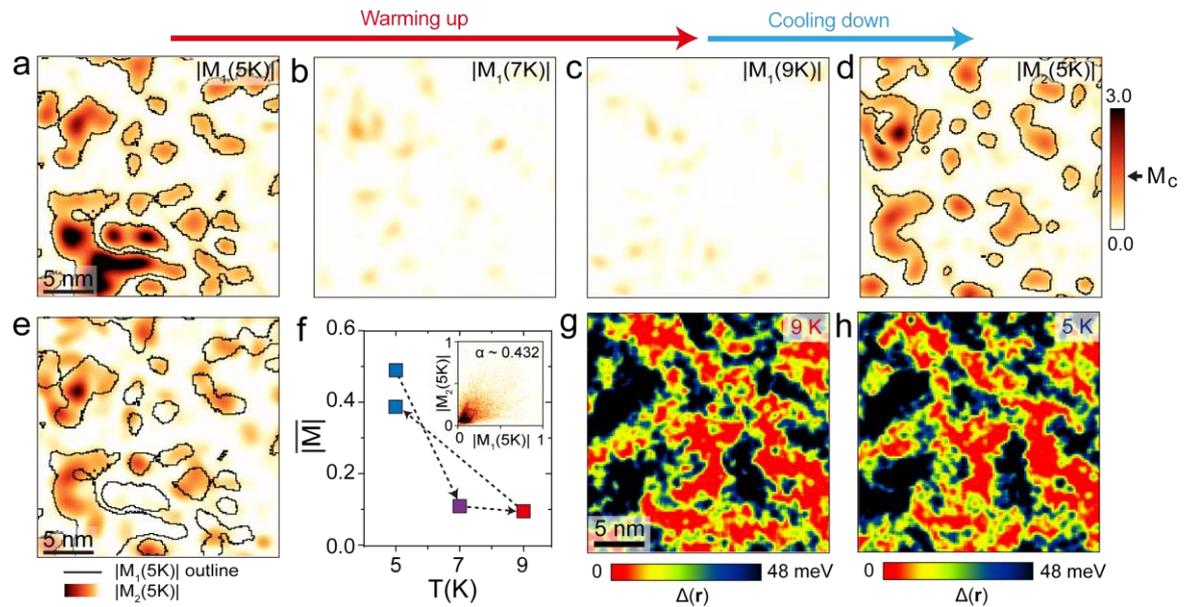

**Figure 4.** Evolution of magnetic and electronic properties of $x$~3.4% La-substituted $Sr_3Ir_2O_7$ along a thermal loop. **(a-d)** AF amplitude maps $|M(\mathbf{r}, T)|$ as a function of temperature, all acquired over the same region under identical conditions. Magnetic textures outlined in (a,d) have been extracted using the same amplitude threshold value $M_c$. **(e)** AF amplitude map $|M_2(\mathbf{r}, 5\text{ K})|$ overlapped with a magnetic texture outline of $|M_1(\mathbf{r}, 5\text{ K})|$ map from **(a)**. **(f)** Plot of the average value of $|M(\mathbf{r}, T)|$ in (a-d). Inset in (f) is a 2D intensity cross-correlation histogram of $|M_1(\mathbf{r}, 5\text{ K})|$ and $|M_2(\mathbf{r}, 5\text{ K})|$. **(g,h)** Maps of the spectral gap ($\Delta(\mathbf{r})$) at 9 K and 5 K, respectively, both obtained over the same area as $|M(\mathbf{r}, T)|$ maps. STM setup condition: M(r) maps: $V_{sample}$ = 600 mV, $I_{set}$ = 100 pA, +/- 4 T, 5 K **(a)**, 7 K **(b)**, 9 K **(c)**, and 5 K **(d)**; $\Delta(\mathbf{r})$ maps: $V_{sample}$= 300 mV, $I_{set}$ = 200 pA, $V_{exc}$ = 6 mV (zero-to-peak), 0 T, 9 K **(g)** and 5 K **(h)**.

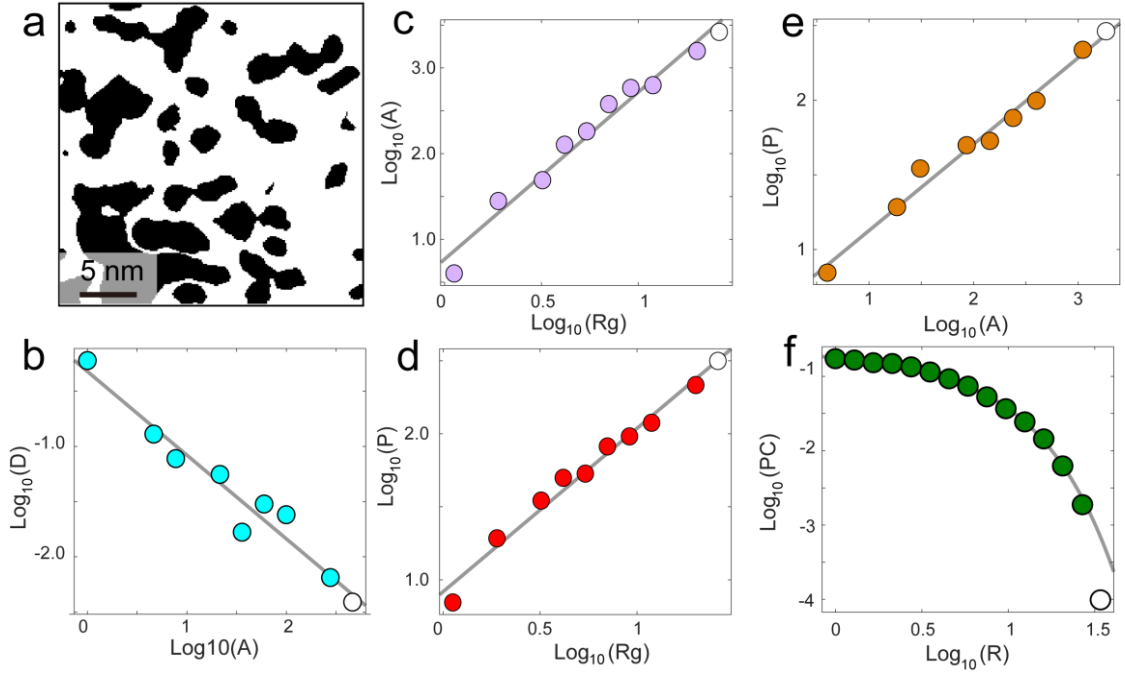

**Figure 5.** Scale-invariant magnetic texture at 5 K. (**a**) Binarized AF domains obtained from $|M_I(\mathbf{r}, 5K)|$ based on an amplitude threshold $M_c$. (**b**) Logarithmically binned domain area distribution $D(A)$, following a scale-free power-law distribution ($D(A) \sim A^{-\tau}$) with the critical exponent $\tau = 0.88 \pm 0.24$. (**c, d**) Area ($A$) and perimeter ($P$) vs. gyration radius ($R_g$) plotted using logarithmic binning. Solid lines are power-law fits of $P \sim R_g^{dh^*}$ and $A \sim R_g^{dv^*}$, with critical exponents $dh^* = 1.20 \pm 0.05$ and $dv^* = 2.0 \pm 0.09$. (**e**) Perimeter ($P$) vs. area ($A$) plot, reflecting the effective fractal dimension ratio $dh^*/dv^*$. The solid line is the power-law fit of $P \sim A^{dh^*/dv^*}$ with $dh^*/dv^* \sim 3/5$. (**f**) Pair connectivity ($PC$) function vs. distance ($r$) plot using logarithmic binning. The solid line is fit to a power-law function with an exponential cutoff $g_{conn} \sim r^{-\eta} \cdot e^{-x/\xi}$, where $\xi$ is the correlation length and $\eta = 0.02 \pm 0.01$ is the exponent for the connectivity function. Values of $P$, $r$, and $R_g$ are in units of pixels and $A$ is in units of area occupied by a single pixel. The single pixel size in the raw STM topograph used to obtain image in (a) is 1.2 Å by 1.2 Å.

**Acknowledgements:**
I.Z. gratefully acknowledges the support from DOE Early Career Award DE-SC0020130. S.D.W., Z.P., and X.C. gratefully acknowledge support from the National Science Foundation via award DMR-1905801. Z.W. acknowledges the support from U.S. Department of Energy, Basic Energy Sciences Grant No. DE-FG02-99ER45747.